\documentclass[epj]{webofc}
\usepackage[varg]{txfonts}   
\usepackage{color}
\usepackage{hyperref}

\renewcommand*{\eqref}[1]{Eq.~(\ref{eq:#1})}

\newcommand*{\figref}[1]{Fig.~(\ref{fig:#1})}
\newcommand*{\figlab}[1]{\label{fig:#1}}

\newcommand*{\seclab}[1]{\label{sec:#1}}
\wocname{\includegraphics[width=0.25cm,clip]{logoARENA} ARENA2018}
\wocname{ARENA2018}
\woctitle{ARENA2018}
\woctitle{\includegraphics[width=0.25cm,clip]{logoARENA} ARENA2018}
\begin{document}
\title{ARIANNA: Current developments and understanding the ice for neutrino detection}
%
%


\author{\firstname{Anna} \lastname{Nelles}\inst{1,2}\fnsep\thanks{\email{anna.nelles@desy.de}} 
       \firstname{for the} \lastname{ARIANNA Collaboration}
}

\institute{Institut f\"ur Physik, Humboldt-Universit\"at zu Berlin, Germany
\and
          DESY, Platanenalle 6, 15738 Zeuthen, Germany
          }

\abstract{%
The ARIANNA experiment aims to detect the radio signals of cosmogenic neutrinos. It is running in its pilot phase on the Ross Ice-shelf, and one station has been installed at South Pole. The ARIANNA concept is based on installing high-gain log periodic dipole antennas close to the surface monitoring the underlying ice for the radio signals following a neutrino interaction. Especially, but not only in this configuration, it is essential to understand the trajectories that the signals take through the ice. We will report on various experimental evidence concerning the signal propagation in ice. We will discuss the implications for neutrino detection, results of neutrino searches and give the first introduction to a new modular simulation framework. 
}
\maketitle
\section{Current status}
\label{saec-status}
The ARIANNA experiment is currently running in its pilot-phase, the Hexagonal-Radio-Array (HRA) on the Ross Ice-Shelf. The main HRA stations have been running reliably since 2014. In the years following the original deployment several special purpose stations to detect cosmic rays have been commissioned (see \cite{Glaser_proc}). The HRA comprises seven stations equipped with four downward-facing log-periodic dipole antennas (LPDAs), one pair aligned north-south and the other pair orthogonally to the first pair. The fully autonomous stations run on solar-panels and batteries. Communications is handled either via the Iridium satellite network or via long-range wifi, connecting via a relay to McMurdo station. 

In 2016 a first custom-made wind turbine was installed that was found fully functional after the Antarctic winter. The recorded power data was promising, albeit not sufficient to power a station. The second generation wind turbine was installed in 2017 and, as of June 2018, it is powering a station during polar night. The batteries charge quickly during periods of high winds, which keeps the station also running after the winds have died down again. A third version of the turbine is currently being constructed, which is further optimized for the windspeeds and temperatures in Moore's Bay. In order to power autonomous stations through the polar night a wind turbine that does not create RF noise and can power a station even at low windspeeds is a prerequisite. 

\section{Neutrino sensitivity and searches}
\seclab{sec-neutr}
After deriving a first limit on the neutrino flux in \cite{Neutrino} with early ARIANNA HRA data, the incoming data is revisited on a regular basis to ensure that the array is still operating according to expectations.  Along the lines of the analysis used to identify cosmic rays \cite{CosmicRays}, the analysis to identify neutrinos was improved to take into consideration both the correlation with neutrino templates as well as the signal amplitude. Cuts were developed that would retain 90\% of the neutrinos. Using data from the season of 2016/17, in most stations no events were found in the signal region. Some stations, however, showed events in the signal region (see \figref{neutrino}). One of the events arrived in coincidence with an event recorded by a dedicated cosmic ray station that is uniquely able to tag cosmic rays, due to its upward facing antennas. Due to the similarity of all events found in the signal region for the neutrino analysis, this suggests that air shower signals can mimic neutrino signals, if the station has no means of distinguishing a signal coming from the air from one coming from the ice below. Since the cosmic ray station does not have an uptime of 100\% and the radio signal of an air shower can extend over as little as $\sim100$ meters, the tagging via special-purpose stations is not enough. This is the reason why production stations for a large radio array must include upward facing antennas to uniquely tag cosmic rays. As cosmic rays are much more abundant than neutrinos, it is essential that all events are identified and removed in neutrino searches. Upward facing antennas, will also provide an additional handle for noise rejection, as noise also likely comes from above. 

\begin{figure*}
\centering
\includegraphics[width=0.8\textwidth]{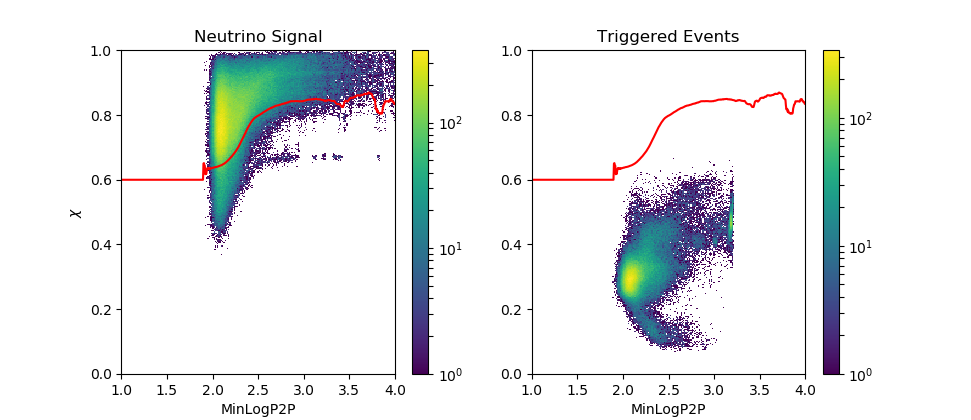}
\includegraphics[width=0.7\textwidth]{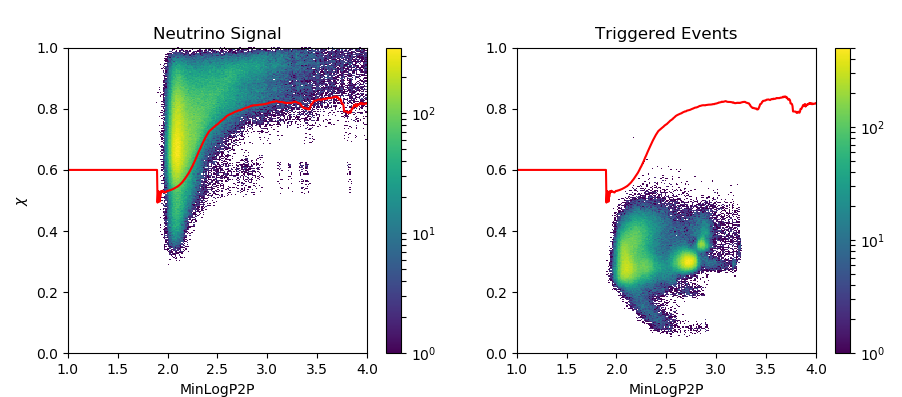}
\caption{Neutrino search for two stations of the HRA (top and bottom row). The measured data are given as distribution of average correlation to the signal templates (y-axis) over the pulse amplitude (x-axis). The red line indicates the cut above which 90\% of the simulated neutrino signal would be retained. The background distributions look slightly different per station with might be related to the depth of the deployed hardware, the positioning of the towers or other unaccounted differences in the station hardware. The lower figure shows an event in the signal region, which has been tagged as cosmic ray by another station.}
\figlab{neutrino}       
\end{figure*}

\section{Ice studies and propagation}
\seclab{sec-Ice}
A strong influence on the sensitivity of radio detectors for neutrinos is caused by the propagation of the emission in ice. Classically a smooth index of refraction profile, 
continuously increasing with depth was assumed that leads to bent trajectories of the emission, which has to be taken into account when reconstructing signals. It also leads to zones that can never be reached by the emission. These are usually referred to as \emph{shadow} zones. In general, the allowed propagation paths strongly influence the effective volumes of neutrino detectors. Deeper stations can profit from signals being bent down to the detector, while stations at the surface lose visible volume for certain geometries. 

In order to test the signal propagation several campaigns with radio pulsers were conducted using ARIANNA stations both in Moore's Bay and at South Pole. 

\subsection{Pulser at South Pole}
In the polar season of 2017/18 a transmitter was lowered into the SPICE hole \cite{SPICE} at South Pole (see \figref{layout}). The pulses triggered both radio neutrino experiments deployed at South Pole, ARA and ARIANNA. We focus here on the ARIANNA data. The schematic set-up is shown in \figref{layout}. Since no ARIANNA personnel was present during the tests, the station ran in its normal data-taking mode, which is occasionally interrupted for communications, which means that some episodes containing pulses could not be recorded.  

During data taking, South Pole station was open and the ARIANNA station subject to a number of noise triggers. Consequently, the pulses originating from the pulser in the SPICE hole had to be selected first. For this a procedure was designed that used typical recorded signal shapes as search templates. All recorded data was classified according to these templates (different types of noise pulses, long and short pulses seemingly stemming from the SPICE hole) and then time differences between events were inspected. As the pulser was set to have a fixed periodicity, this should allow for a clean signal identification. However, it seems that the pulser was not stable when being lowered, so it started emitting two types of pulses at a non-constant frequency, which complicates the search especially in the forbidden region, where the pulse quality observed was less clean. 

Selecting one type of pulse, which appeared reasonably stable, allows for the reconstruction of the signal direction as shown in \figref{hole_data}. In the nominally allowed signal zone, the position of the pulser can be well reconstructed with a sub-degree resolution. However, a systematic absolute offset has been observed. This could be explained due to the fact that the station geometry is not known to the precision needed for the reconstruction. This will have to be taken into consideration when deploying a large array, by developing a precision positioning system for the antennas or a calibration instrument. 

\begin{figure*}
\centering
\includegraphics[width=0.25\textwidth]{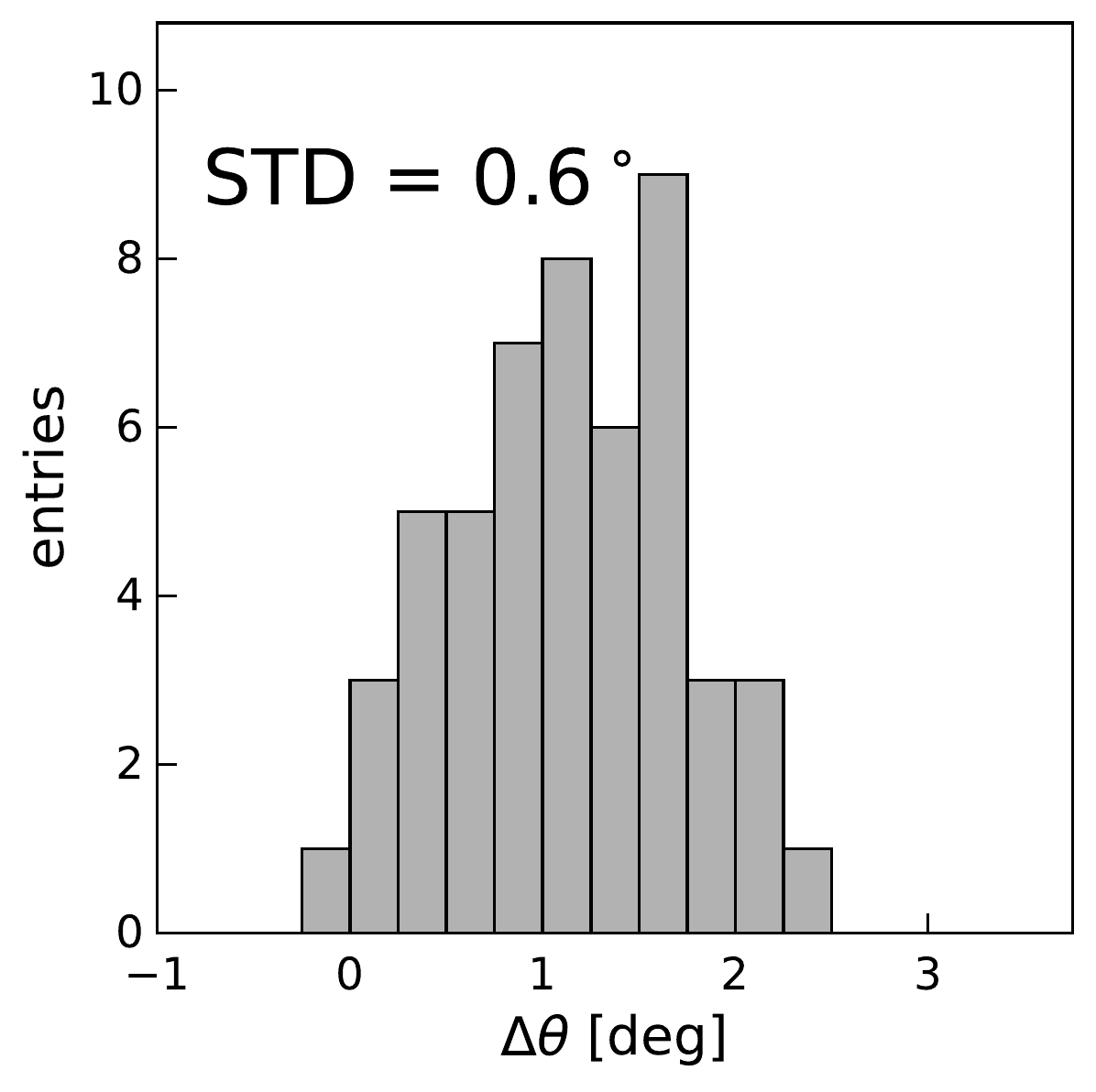}
\includegraphics[width=0.45\textwidth]{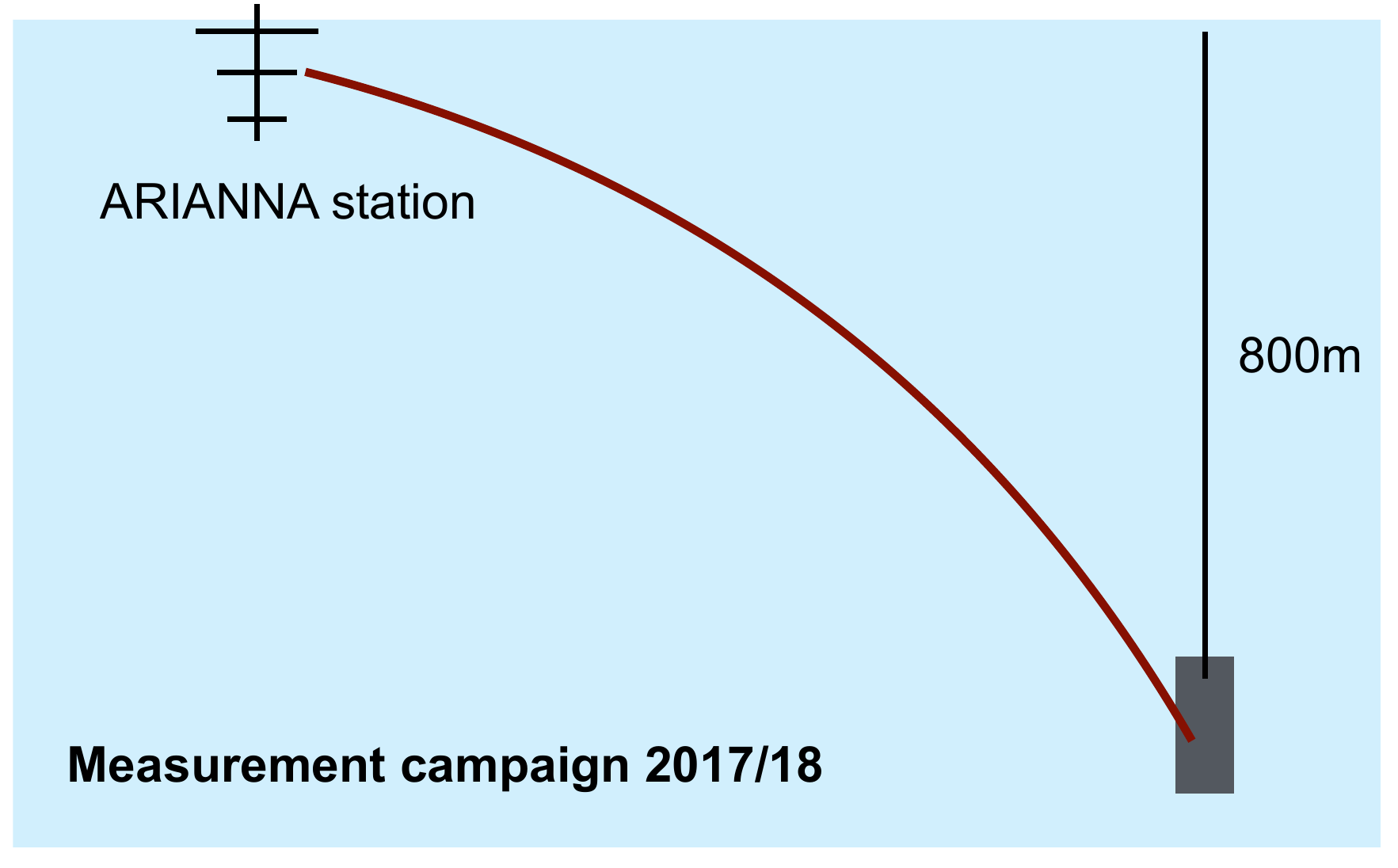}
\includegraphics[width=0.25\textwidth]{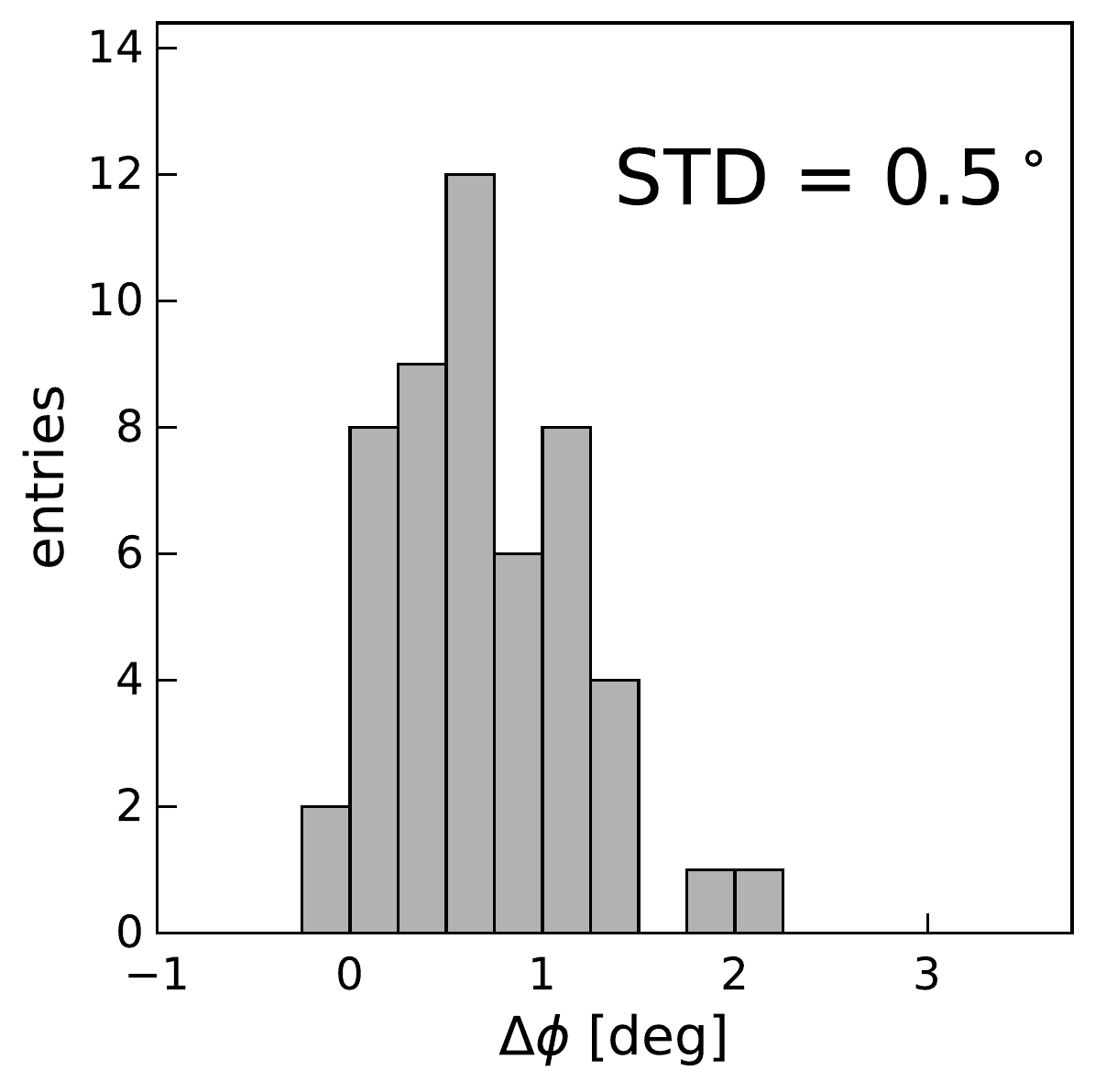}
\caption{Set-up of pulser lowered into the SPICE hole at South Pole in season 2017/18. The middle figure shows the schematic set-up, the signal path is an illustration. The pulser was lowered continuously. The left and right histogram show the resolution achieved in zenith and azimuth angle when reconstructing this pulser run.}
\figlab{layout}       
\end{figure*}

\begin{figure*}
\centering
\includegraphics[width=0.45\textwidth]{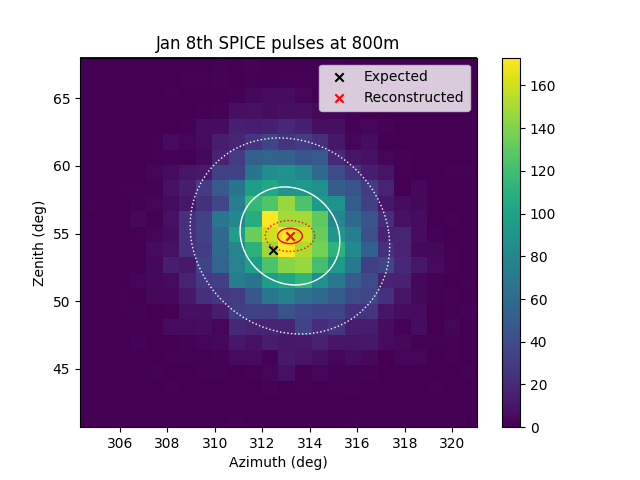}
\includegraphics[width=0.45\textwidth]{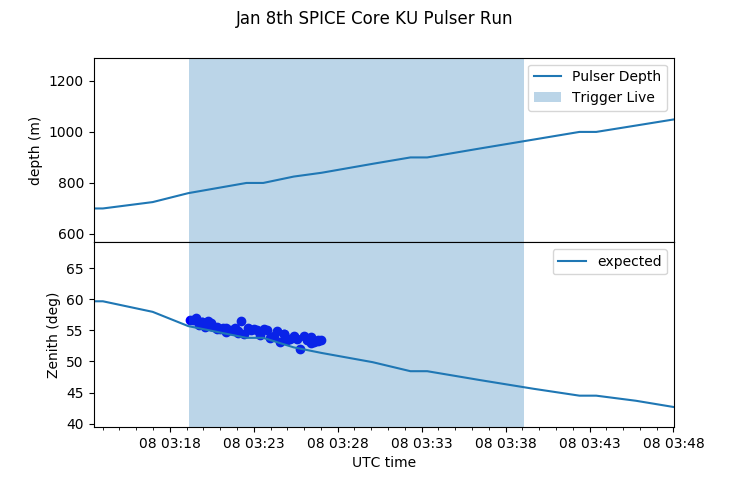}
\caption{More preliminary geometric results from the SPICE hole tests. Left: Absolute pointing of the reconstruction. The red lines indicate the statistical uncertainty, while the white lines show the systmatical uncertainty (1 and 2 $\sigma$ respectively). The station geometry is only known to about 0.3 meter per antenna, which is the driving systematic uncertainty on reconstruction of the absolute position of the pulser. Right: Depth of the pulser as function of time. The reconstructed zenith angle is also shown. The ARIANNA station at South Pole was recording data in a window of 20 minutes (blue). It has been observed that the pulser lost stability due to the high pressure in the hole. The test will be repeated in forthcoming seasons. }
\figlab{hole_data}       
\end{figure*}

\subsection{Propagation in the shadow zone}
Recent measurements at South Pole, Moore's Bay \cite{Ice_paper} and Greenland \cite{Greenland_ice} have shown that the signals can be observed in the shadow zones despite predictions of the contrary. Signals from a transmitter in a hole of 19 meters were observed in all operational ARIANNA stations at distances of up to 2 km. Also for the South Pole studies, strong signals were observed in the nominally shadowed zone. However, the signal characteristics were less well-defined. Also due to the problematic pulser behavior, reconstructing these data is on-going work. Still, preliminary results show that some fraction of the recorded events has been reconstructed to match the pulser position in the shadow zone, as shown in \figref{hole_data}.

\subsection{Observation of two pulses}
When considering the signal propagation in ice of an antenna not so far below the surface, it is likely that the antenna can record both the direct signal of a neutrino and the signal reflected from the surface. The time-delay between those two signals will carry information about the distance to the interaction vertex and therefore will provide a way to improve the energy reconstruction. The data that allow for the testing of this hypothesis stem from one of the special purpose stations built in the HRA to detect cosmic rays from the horizon. The LPDAs are elevated above the ice surface and therefore record both the direct signal, as well as the signal reflected off the surface. One example event is shown in \figref{double_pulse}. After correcting for the antenna and amplifier response, two pulses with different polarity are visible. Their distance in time corresponds to what is expected from the reconstructed arrival direction of the air shower. This illustrates that a double-pulse effect is already resolvable at a distance of only 5 meters from the reflecting surface with the current ARIANNA system. 

\begin{figure*}
\centering
\includegraphics[width=0.45\textwidth]{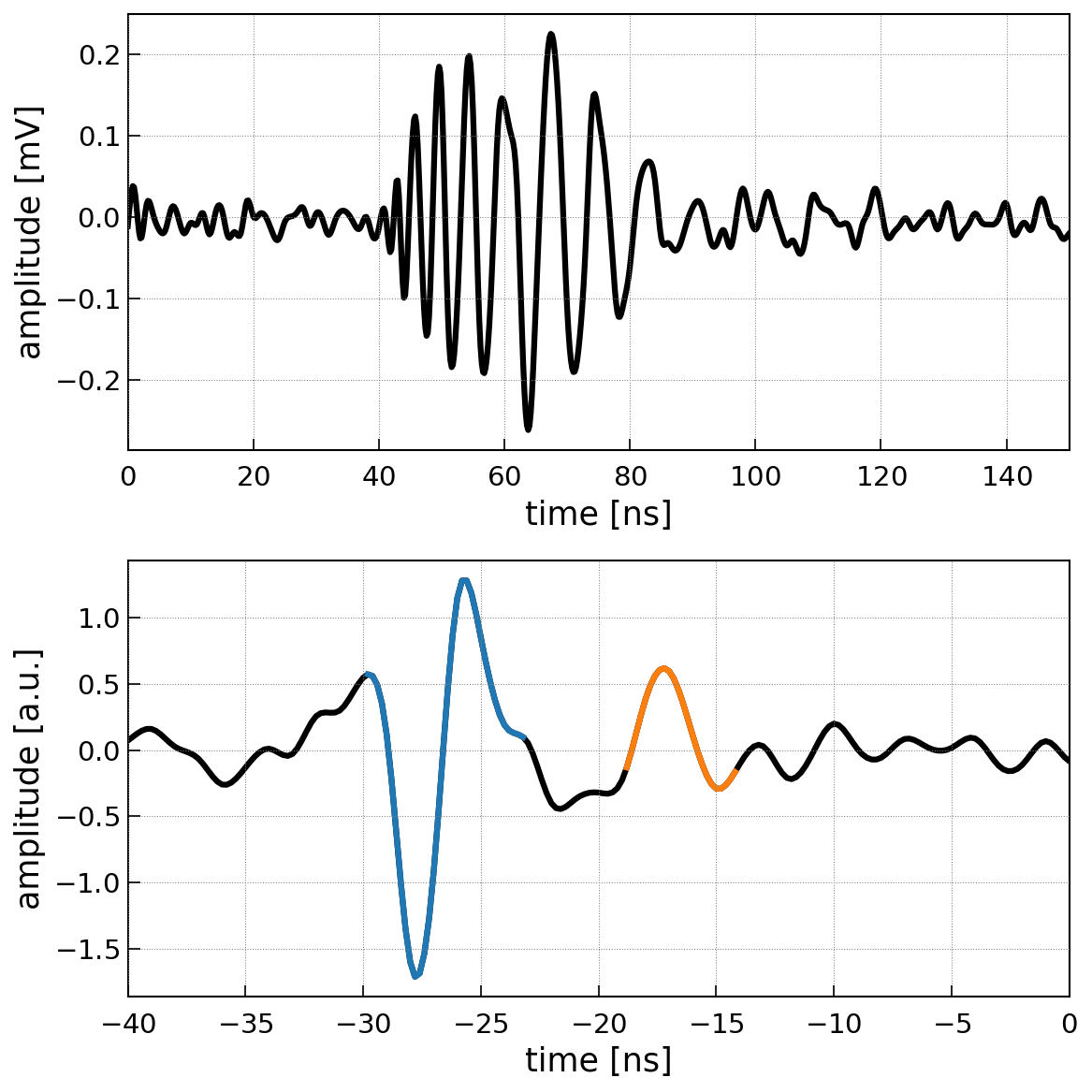}
\includegraphics[width=0.45\textwidth]{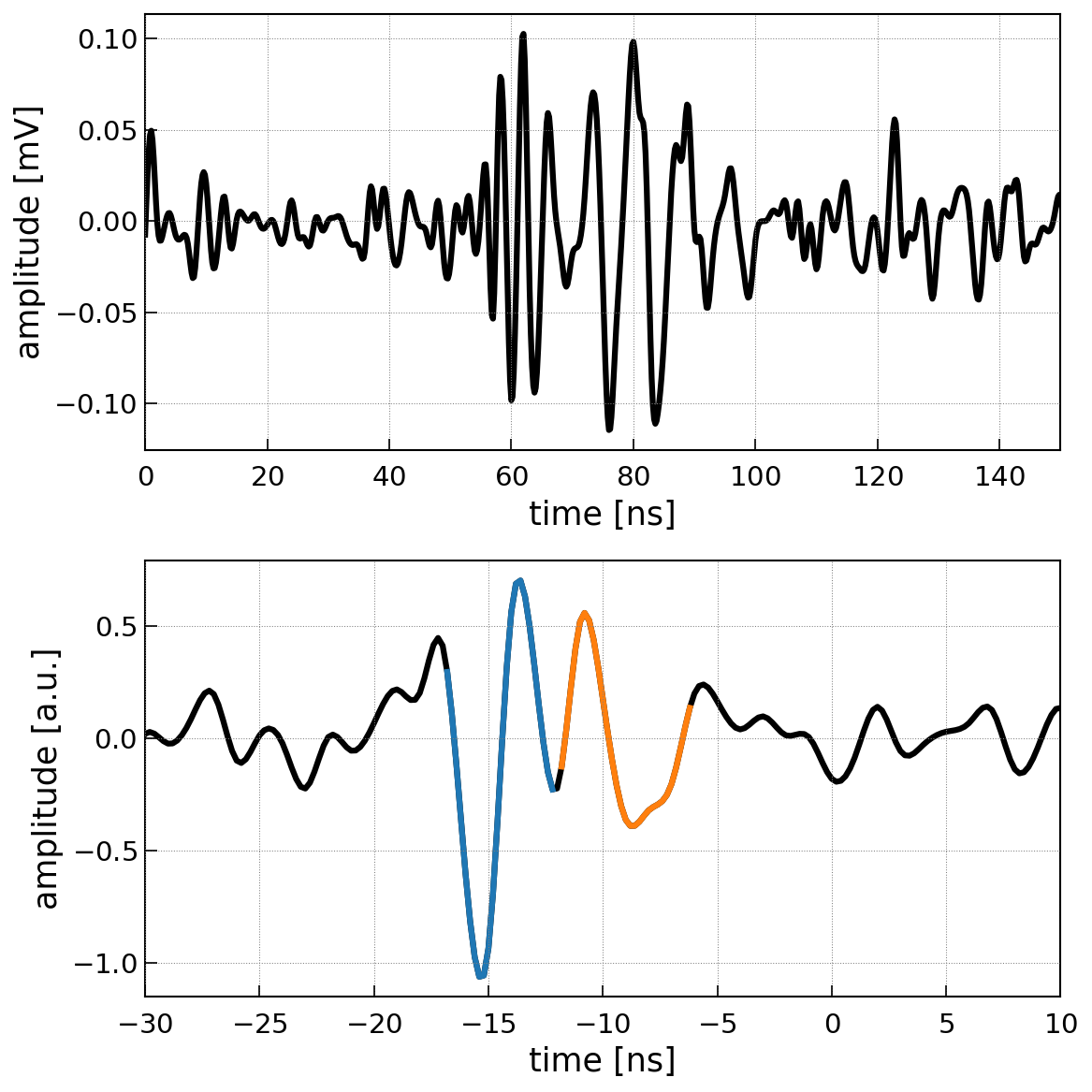}
\caption{Air shower signals as measured in an LPDA elevated 5 meters above the Ross-Ice Shelf. Two events are shown. The one on the left has an arrival zenith angle of $69^{\circ}$, the one on the right $81^{\circ}$ . In the top row, the raw data is shown, while the bottom row shows the data after correcting for the antenna sensitivity and amplifiers. Both the direct signal and the reflected signal become clearly distinguishable after the unfolding of the hardware. The detecting antenna was rotated as such that it measured the horizontally polarized component of the emission, which is strongest as the magnetic field is pointing almost straight up. }
\figlab{double_pulse}       
\end{figure*}

\section{A new Monte Carlo Framework}
The increased complexity of the signal propagation and the desire to design an optimal detector is pushing the limits of the current signal Monte Carlo programs. All programs currently in use by experiments (ARASim for ARA, ShelfMC for ARIANNA and the ANITA simulation framework) are tailor-made to their use case and not flexible enough to easily include different propagation models, different emission models or detector designs. Following a first effort of comparing Monte Carlo codes \cite{Tevpa}, a new modular Monte Carlo Framework has been started (\figref{MC}). The project is open to anyone\footnote{https://github.com/nu-radio/NuRadioMC} and aims to allow for the inclusion of all experimental efforts relating to the radio detection of neutrinos. In a modular approach, clear interfaces between the four pillars are defined, which will allow for the flexible interchange of different components. Interfaces to air shower simulations such as CoREAS are also foreseen. 

\begin{figure*}
\centering
\includegraphics[width=0.7\textwidth]{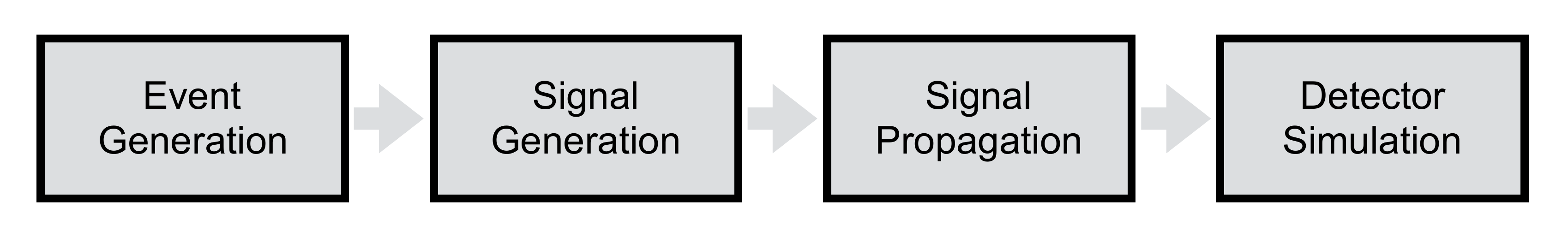}
\caption{Schematic set-up of the simulation framework and its four pillars. The event generation module will pass neutrino properties to the signal generation, from where the electric field at the source is passed to the signal propagation module, which calculates the electric field at the detector. The detector simulation is flexible so that it can account for various types of hardware and array configurations.}
\figlab{MC}       
\end{figure*}

\section{Conclusions and future plans}
The ARIANNA HRA\footnote{The ARIANNA HRA is supported by the National Science Foundation. AN is supported by the German Research Foundation, (DFG), on grant NE 2031/2-1. } has been running stably on the Ross Ice-Shelf. The search for neutrino signals is performing according to expectations and the collaboration is focussing on hardware developments such as a new wind turbine to increase life time during the polar night and improving the understanding of real polar ice. To that end, several measurement campaigns have been conducted, studying the signal propagation. In contrast to classical expectations, signals have been found to propagate over kilometers at the surface, which might provide an opportunity for future radio neutrino detectors. 

The next step in developing the radio detection of neutrinos in ice, has to be the evaluation of the experiences of all current experiments and the development of a common way forward toward a large science grade instrument. One way forward is illustrated in \cite{Barwick_proc}.

%
%
%

\end{document}